\documentclass[aps,prl,preprint,superscriptaddress]{revtex4}
\usepackage{dcolumn}	
\usepackage{bm}			
\usepackage{here}
\usepackage[dvipdfm]{}
\usepackage{graphicx}
\usepackage{epstopdf}
\usepackage{mathrsfs}
\usepackage{amsmath}
\def\tc{$T_{\rm c}$}

\def\cx{Cu$_x$Bi$_2$Se$_3$}
\def\sx{Sr$_x$Bi$_2$Se$_3$}

\def\BS{Bi$_2$Se$_3$}
\def\hct{$H_{\rm c2}$}
\def\dv{\textbf {d}-vector}
\begin{document}

\title{Direction and symmetry transition of the vector order parameter in topological superconductors Cu$_x$Bi$_2$Se$_3$
}
\author{T. Kawai}
\thanks{These authors contributed equally.}
\affiliation{Department of Physics, Okayama University, Okayama 700-8530, Japan}
\author{C. G. Wang}
\thanks{These authors contributed equally.}
\affiliation{Institute of Physics, Chinese Academy of Sciences, and Beijing National Laboratory for Condensed Matter Physics,  Beijing 100190, China}
\affiliation{School of Physical Sciences, University of Chinese Academy of Sciences, Beijing 100190, China}
\author{Y. Kandori}
\affiliation{Department of Physics, Okayama University, Okayama 700-8530, Japan}
\author{Y. Honoki}
\affiliation{Department of Physics, Okayama University, Okayama 700-8530, Japan}
\author{K. Matano}
\affiliation{Department of Physics, Okayama University, Okayama 700-8530, Japan}
\author{T. Kambe}
\affiliation{Department of Physics, Okayama University, Okayama 700-8530, Japan}
\author{Guo-qing Zheng}
\thanks{To whom correspondence should be addressed; E-mail:  gqzheng123@gmail.com}
\affiliation{Department of Physics, Okayama University, Okayama 700-8530, Japan}
\affiliation{Institute of Physics, Chinese Academy of Sciences, and Beijing National Laboratory for Condensed Matter Physics,  Beijing 100190, China}

\begin{abstract}
	Topological superconductors have attracted wide-spreading interests  for the bright application perspectives   to quantum computing. Cu$_{0.3}$Bi$_2$Se$_3$ is a rare  bulk topological superconductor with an odd-parity wave function, but
	the details of the vector order parameter $\textbf{{d}}$ and its pinning mechanism are still unclear. 
    We have succeeded in growing Cu$_x$Bi$_2$Se$_3$ single crystals with unprecedented high doping levels. 
	For samples with $x$ = 0.28, 0.36 and 0.37 with similar carrier density as evidenced by 
	 Knight shift, the in-plane upper critical field
$H_{\rm c2}$ shows  a two-fold symmetry. However,
 the angle at which the $H_{\rm c2}$ becomes minimal is different by 90$^\circ$ among them, 
	which  indicates  that the $\textbf{{d}}$-vector direction is different for each crystal 
	 likely
	  due to a different local environment. The carrier density for
	   $x$ = 0.46 and 0.54 increases substantially compared to $x\leq$  0.37. 
	   Surprisingly, the in-plane $H_{\rm c2}$ anisotropy disappears, indicating that the gap symmetry 
	undergoes a transition from  nematic to  isotropic (possibly chiral) as carrier increases. 
\end{abstract}

\maketitle
Exploring topological materials  and  their electronic functions are among the front-most topics of  current condensed matter physics.
In particular,   much attention has been paid in recent years to topological superconductors where
Majorana fermions (excitations) are expected to appear on edges or in the vortex cores\cite{QiZhang,Sato_Ando_review_2017}. 
Such novel edge states  can  potentially  be applied  to  fault tolerant
non-Abelian quantum computing \cite{TopologicalQuantumComputation,quantumcomputer_KITAEV20032}.
So far, great success has been achieved in observing the Majorana bound state  on the interface of a ferromagnet or a topological insulator in proximity to an $s$-wave superconductor\cite{Fu_PhysRevLett.100.096407,Kouwen,Yazdani,Jia},
or on the surface of iron-based superconductors \cite{Ding}. In contrast, research on bulk topological superconductors progresses much more slowly. 
Candidates of bulk topological superconductors include 
superconductors with broken time reversal symmetry \cite{Senthil,Wilczek}, superconductors with broken spatial inversion symmetry \cite{Nishiyama,sato_PhysRevB.79.094504} 
and  odd-parity superconductors with spatial inversion symmetry\cite{Fu_Berg_PhysRevLett.105.097001,sato_odd_paritySC_PhysRevB.81.220504}.
For the last case, the criteria for  topological superconductivity 
are effectively two fold. Namely, odd-parity of the gap function and an odd number of time-reversal invariant momenta in the Brillouin zone\cite{Fu_Berg_PhysRevLett.105.097001}.
Experimentally, clear evidence for odd-parity superconductivity had not been found until very recently
\cite{MatanoKrienerSegawaEtAl2016}.  
Although Cu-doped topological insulator  \cx 
\cite{Hor_PhysRevLett.104.057001} 
 had been proposed as a candidate \cite{Fu_Berg_PhysRevLett.105.097001}, 
 experiments had been controversial\cite{Sasaki_PhysRevLett.107.217001,Peng_NoZBCP_PhysRevB.88.024515,STM_PhysRevLett.110.117001}.

The discovery of spontaneous spin rotation-symmetry breaking in the bulk superconducting state of Cu$_{0.3}$Bi$_2$Se$_3$ by nuclear magnetic resonance (NMR) measurements established the spin-triplet, odd-parity superconducting state \cite{MatanoKrienerSegawaEtAl2016}.
Since there is only one time-reversal-symmetric momentum in the  Brillouin zone of \cx\ \cite{ARPES}, this material fulfills the two-fold criteria and can be classified as a  topological superconductor. 
However, detailed gap function is still unclear.
If the gap is fully-opened, then  Cu$_{0.3}$Bi$_2$Se$_3$ is  a   class DIII topological superconductor \cite{class}, where Majorana zero-energy modes are expected at edges or vortex cores.
If there are nodes in the gap function, the material is nonetheless topological
just as the cases of  Dirac or Weyl semimetals.

A more generally-used term associated with the gap in 
 a spin-triplet superconductor is the vector order parameter \textbf{{d}},  
 whose direction is perpendicular to the direction of paired  spins  and whose magnitude is the gap size. The \dv\ 
  was  found to be parallel  to  $a$-axis (the Se-Se bond direction) in Cu$_{0.3}$Bi$_2$Se$_3$ \cite{MatanoKrienerSegawaEtAl2016}.   This was the first case where the \dv\ direction was unambiguously determined in any spin-triplet superconductor candidate. 
The emergent two-fold symmetry in the Knight shift below $T_{\rm c}$  \cite{MatanoKrienerSegawaEtAl2016} was interpreted by the concept of nematic order \cite{Fu_CuxBi2Se3_PhysRevB.90.100509}, and  had triggered many subsequent extensive studies on rotational symmetry breaking 
by various methods, which also revealed a two-fold symmetry in other physical properties
\cite{Yonezawa_Natphys,Sr_dope_2fold_PanNikitinAraiziEtAl2016,Asaba_Sr_PhysRevX.7.011009,Sr_dope_2fold_Du2017,Nikitin_PhysRevB.94.144516}. Measurements by transport\cite{Smylie_Nb_PhysRevB.94.180510}, penetration depth\cite{Smylie_Nb_PhysRevB.96.115145},   and scanning tunneling microscope (STM) \cite{Nb_dope_STM_PhysRevB.98.094523,FengDL} 
 suggesting unconventional superconductivity have also been reported since then.

However, why  the \dv\ 
is  oriented to one of  $a$-axes, and why it is robust against heat cycle through the superconducting transition, remain unknown.
Note that there are  three equivalent $a$-axis directions.
This issue is important as the gap symmetry ("nematicity" indicator) is closely tied to the direction of the \dv\ \cite{Fu_CuxBi2Se3_PhysRevB.90.100509}.
From  material point of view, it had been unclear whether the carrier density can be controlled and how the physical properties would change with changing carrier density. A previous report showed that the Hall coefficient does not change even though the nominal $x$ increases from 0.15 to 0.45 \cite{Kriener_PhysRevB.86.180505}. 
These are the issues we wish to address in this article. 


In this work, we synthesized  Cu-doped \BS\ single crystals by the electrochemical intercalating method.  
Through the measurements of the Knight shift, we find unprecedentedly that the carrier concentration further increases  with increasing $x$ beyond $x$=0.37. 
We  study the angle dependence 
of the upper critical field 
$H_{\rm c2}$ in different crystals. 
For an odd-parity  gap function, the gap anisotropy 
will lead to an anisotropic  $H_{\rm c2}$.  
One then can  obtain knowledge about how the superconducting gap evolves with $x$ by measuring the  $H_{\rm c2}$ anisotropy. 
For samples with  $x$=0.28, 0.36 and 0.37 which have the same size of the Knight shift, we find a two-fold symmetry in the in-plane $H_{\rm c2}$ 
 by ac susceptibility and magnetoresistance measurements, in agreement with previous reports \cite{MatanoKrienerSegawaEtAl2016,Yonezawa_Natphys,Asaba_Sr_PhysRevX.7.011009,Sr_dope_2fold_Du2017,Sr_dope_2fold_PanNikitinAraiziEtAl2016,Nikitin_PhysRevB.94.144516}. 
However, the angle at which $H_{\rm c2}$ is a minimum  differs by 90$^\circ$, 
which means that the direction of the \dv\ is different for each crystal.
In contrast, for  $x$=0.46 and 0.54, two-fold anisotropy disappears, which indicates a nematic-to-isotropic  transition of the gap symmetry as carrier density  increases.  We discuss possible exotic (chiral) superconducting state for the samples with large $x$.


\section{Results}
\textbf{Sample characterization.} 
Figure 1 shows the result of dc susceptibility measurements for   representative samples.
In Fig. 2, we summarize the properties  of all the samples we synthesized.
The obtained \tc\ and  shielding fraction (SF) for most samples are close to the values reported by Kriener et al.\cite{Kriener_PhysRevB.84.054513}. The SF for  $x$=0.46 is the highest (56.2\%) among those reported so far.
In Table 1 we list the properties for the five samples that we will discuss in this paper. 
\begin{figure}[htbp]
	\includegraphics[clip,width=60mm,pagebox=cropbox]{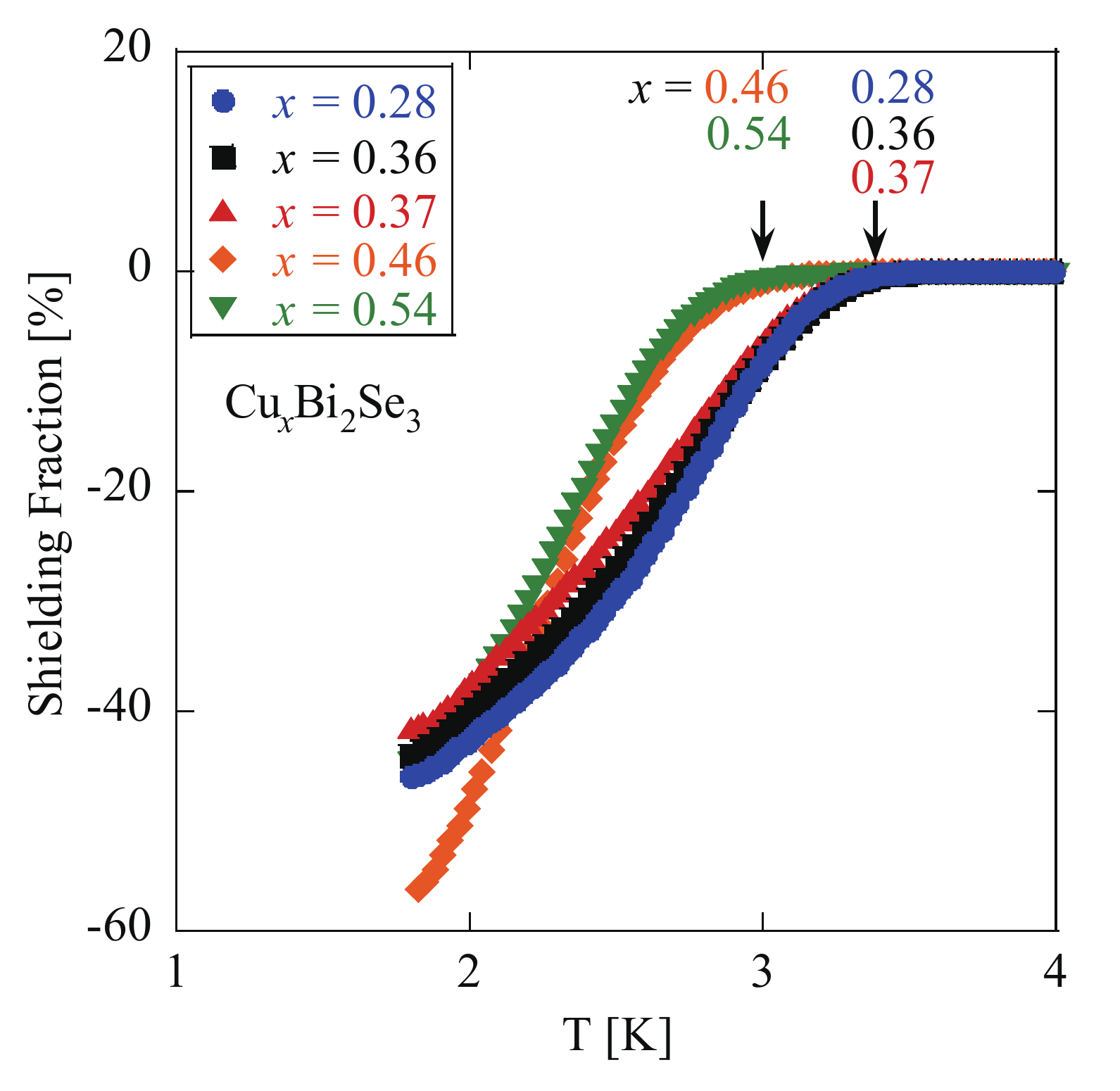}
	\caption{\label{dcchi}
 Diamagnetism measurements. Shielding fraction for five  samples with different $x$. The arrows indicate $T_c$.
	}
\end{figure}

\begin{figure}[htbp]
	\includegraphics[clip,width=65mm,pagebox=cropbox]{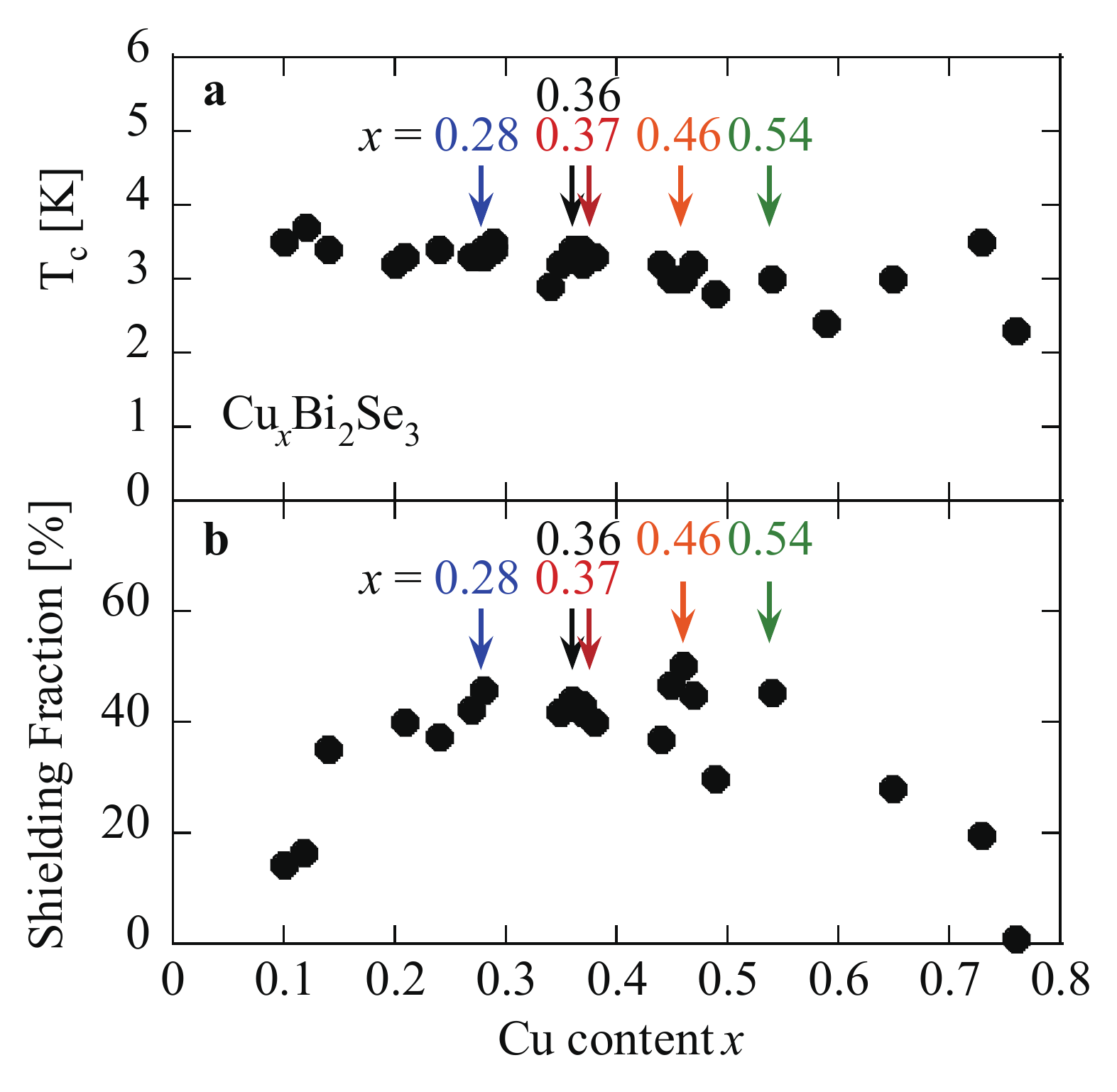}
	\caption{\label{dcchi}
		Properties of the samples.
		Cu content ($x$) dependence of {\bf a}.
		the superconducting critical temperature $T_c$ and {\bf b}. the shielding fraction at 1.8 K.
		The arrows indicate  samples used for the \hct\ 
		and NMR measurements reported in this paper.
	}
\end{figure}
\begin{table}[h]
	\centering
 \begin{tabular}{cccccc}
	\hline\hline
	Cu content $x$ &0.28&0.36&0.37&0.46&0.54\\
	\hline
	\tc [K]& 3.4 & 3.4 & 3.4 & 3.0&3.0 \\
	Shielding Fraction [\%]&45.9 &44.1 &41.5&56.2&44.7 \\
	\hline\hline
\end{tabular} 
	\caption{\label{tab:5/tc}The  \tc\ 
		and the shielding fraction at $T$= 1.8 K for the five   Cu contents discussed in this paper.}
\end{table} 
\begin{figure}[htbp]
	\includegraphics[clip,width=80mm,pagebox=cropbox]{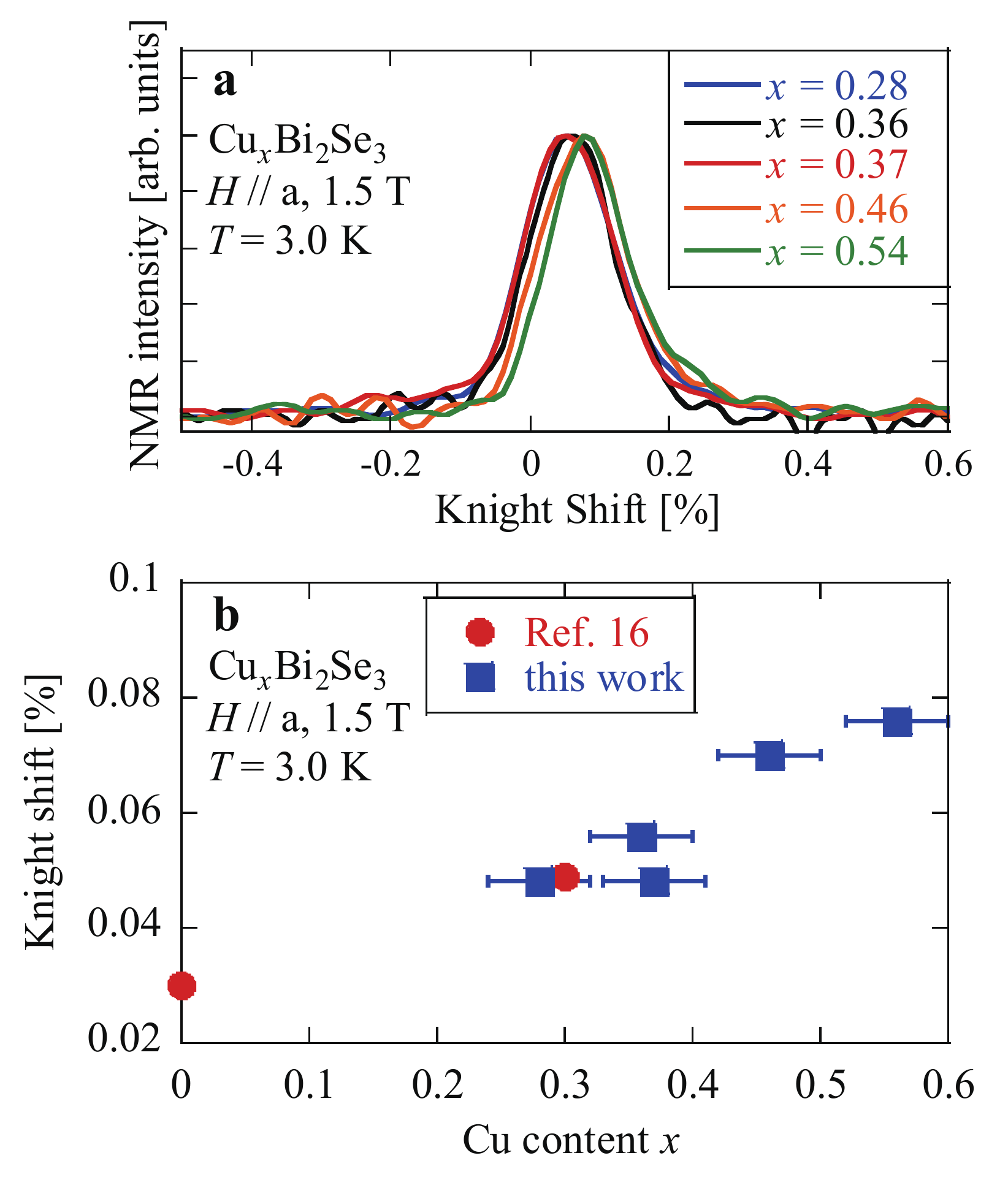}
	\caption{\label{Kvsx}
		NMR Spectra and Knight shift.
	{\bf a}. the $^{77}$Se-NMR spectra measured in a magnetic field of $H_0$ = 1.5 T.
	{\bf b}. Cu content $x$ dependence of the Knight shift. Data for $x$=0 and 0.3 are from Ref. \cite{MatanoKrienerSegawaEtAl2016}.
	The error for Cu-content $x$  was determined by the resolution of the electronic balance used for measuring the weight of Cu-wire.
    }
\end{figure}
Figure \ref{Kvsx}a shows the $^{77}$Se($I = 1/2$)-NMR spectra above \tc\ for five samples with different $x$.
Each spectrum can be fitted by a single Gaussian.
The Knight shift  is $\sim$0.048 \% for $x$ = 0.28, 0.36 and 0.37, 
which is close to the value ($K$ = 0.049 \%) previously reported  for the sample with $x$ = 0.3 \cite{MatanoKrienerSegawaEtAl2016}.
On the other hand,  the Knight shift increased substantially in the samples with  $x$= 0.46 and 0.54.
Generally, the Knight shift is expressed as,
\begin{eqnarray}
K &&= K_{\rm orb}+K_{\rm s} \\
K_{\rm s} &&= A_{\rm hf}\chi_{\rm s},
\end{eqnarray}
where $K_{\rm orb}$ is the contribution due to orbital susceptibility and 
$A_{\rm hf}$ is the hyperfine coupling constant,  which are independent of carrier density. The $\chi_{\rm s}$ is the spin susceptibility which is proportional to electronic  density of states.
For this field configuration, $K_{\rm orb}$=0.03 \%\cite{MatanoKrienerSegawaEtAl2016}.
In Fig. \ref{Kvsx}b, we plotted the Knight shift as a function of Cu content $x$. Upon doping, $K$ increases as compared to the parent compound, which implies that the carrier  is indeed doped into the sample.
The fact that $K$ has a similar value for the  samples with $x$ = 0.28 $\sim$ 0.37 indicates that the carrier density does not change appreciably in  this $x$ region.
Such behavior is  consistent with other reports by different methods\cite{Kriener_PhysRevB.86.180505,Sr_position_PhysRevMaterials.2.014201}.
On the other hand, in the high-$x$  region, we discovered for the first time that carriers increase further as $x$ increases beyond 0.37.

We briefly comment on 
possible mechanisms of Cu-doping. 
In 
Sr-doped system  \cite{Sr_position_PhysRevMaterials.2.014201}, Sr   goes into multiple sites, namely, the intercalated site in-between the quintuple layer blocks, interstitial site, or  the in-plane site  substituting for Bi. 
Cu(Sr) residing on the intercalated or interstitial site contributes electrons, while Cu(Sr) going to the substituting site contributes holes \cite{Hor_PhysRevLett.104.057001}.
 The stage-like $x$-dependence of $K$ suggests that Cu doping mechanism in each $x$-range is not simple, but rather Cu may also go into multiple sites. 
As post annealing is necessary after electrochemical process in the present case, we speculate that Cu may migrate   into a different site after annealing, 
resulting in the peculiar behavior of  $K$ with respect to $x$.  The  Cu position is one of the  issues that needs to be addressed in future works.

\begin{figure}[htbp]
	\includegraphics[clip,width=70mm]{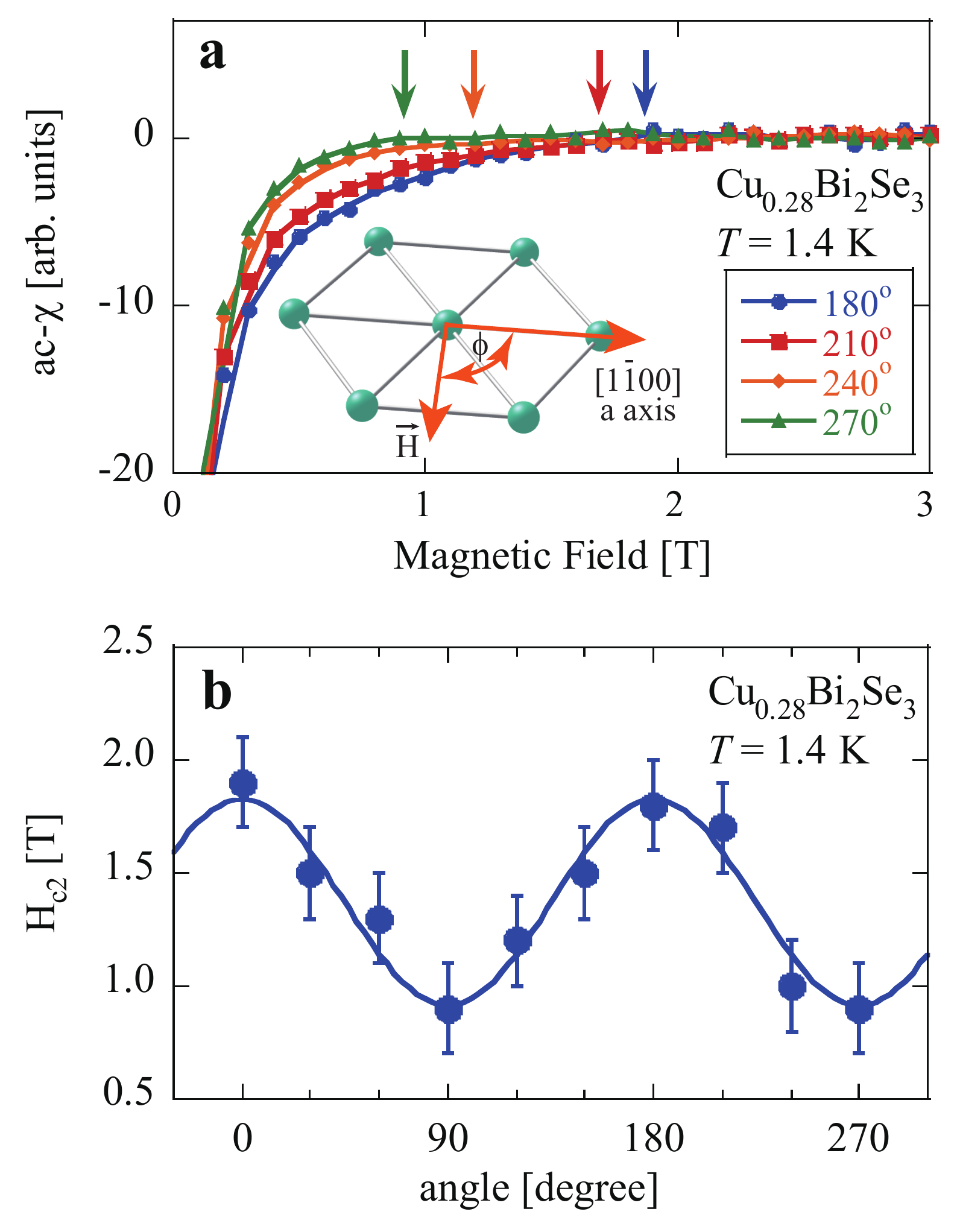}
	\caption{\label{acchi}
		In-plane upper critical field anisotropy determined by magnetic susceptibility.
		{\bf a}, Magnetic field dependence of the ac-$\chi$ at different in-plane angle $\phi$ for the sample with $x$=0.28.  
		The solid curves 
		are guides to the eyes. The arrows  indicate \hct\ for each angle,
		which is defined as a point off the straight line drawn from high-field value (the normal state). The error  was estimated by assuming that the uncertainty equals twice point-interval in measuring  the ac-$\chi$ {\it vs} field curve. For more details see Supplementary Information.
		The	inset 
		is an illustration depicting the hexagonal plane  and  the angle $\phi$.  {\bf b},   \hct\  plotted as a function of in-plane angle $\phi$. The blue curve is  a sine function.  
	}
\end{figure}

\textbf{Anisotropy of the upper critical field \hct\ and its disappearance.}
Next we present data on 
the anisotropy of \hct\ . 
For $x$=0.28, we measured both ac susceptibility  (ac-$\chi$) and electrical resistance  by changing the magnetic field for each field-direction relative to $a$-axis.
The inset to Fig.  \ref{acchi}a shows the angle $\phi$ between the applied magnetic field $H$ and the $a$-axis.
Before Cu doping, the direction of  $a$-axis (Se-Se bond direction) was determined by Laue diffraction, which 
we assign as $\phi$=0 degree. 
The main panel of Fig. \ref{acchi}a shows the magnetic field dependence of the ac-$\chi$ at different  $\phi$ for $x$ = 0.28.
As can be seen in the figure, the field dependence of ac-$\chi$ is clearly different for each angle below a certain field ( \hct\ ), while 
above \hct\, the $H$-dependence of ac-$\chi$ is the same for all angles.
Fig.\ref{acchi}b shows \hct\ as a function of angle. A clear two-fold symmetry is observed.

%
\begin{figure}[htbp]
	\includegraphics[clip,width=80mm]{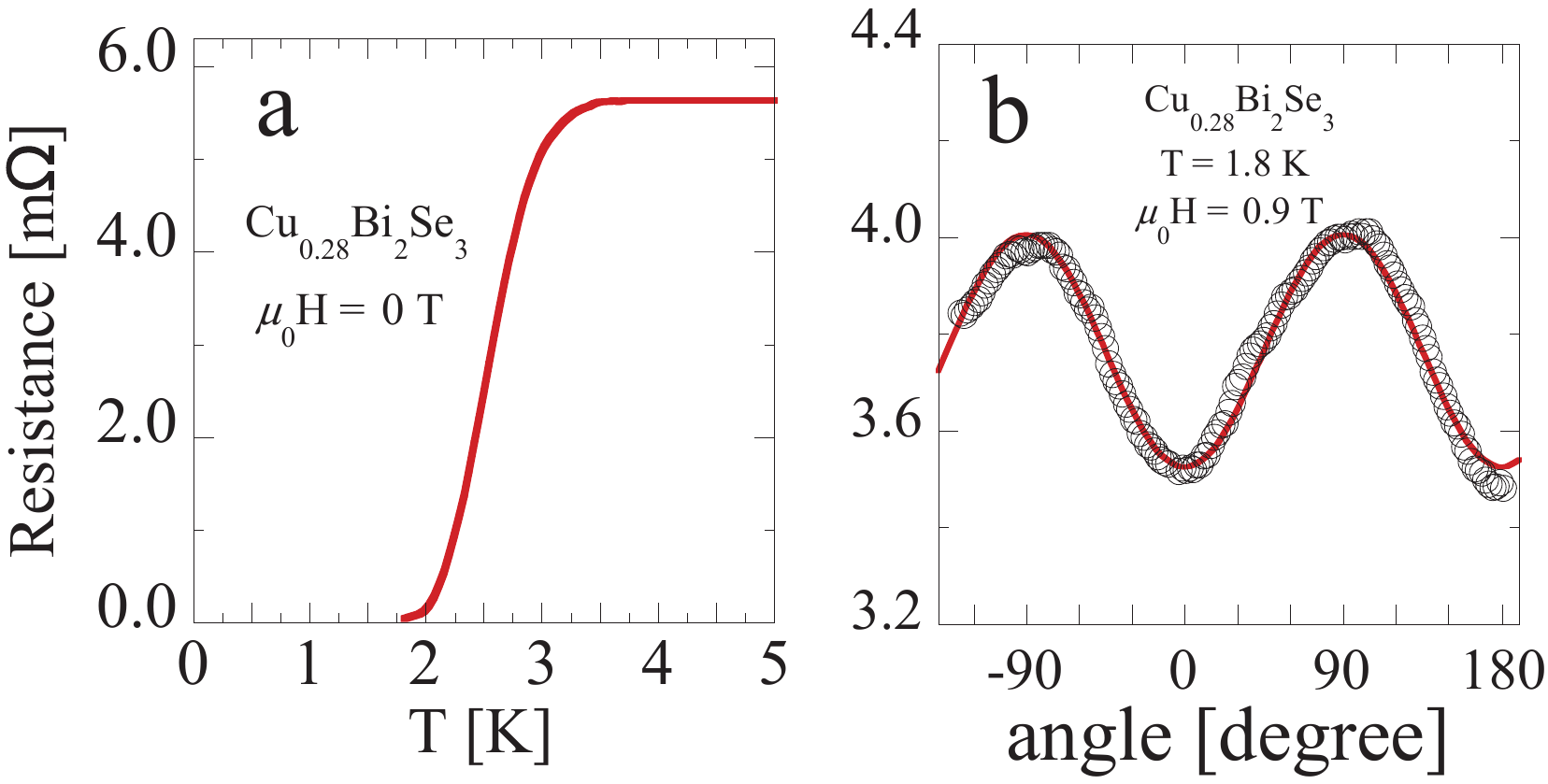}
	\caption{\label{RvsAngle}
		Magneto-resistance measurements.
		{\bf a}. Temperature dependence of the electrical resistance at zero magnetic field, which is in agreement with Ref.\cite{Kriener_PhysRevB.84.054513}.
		{\bf b}. Electrical resistance under a magnetic field of 0.9 T and at $T$=1.8 K as a function of in-plane angle $\phi$. Note that a minimum in the electrical resistance corresponds to a maximal \hct\ . The red  curve is a sine function.
	}
\end{figure}
\begin{figure}[htbp]
	\includegraphics[clip,width=70mm]{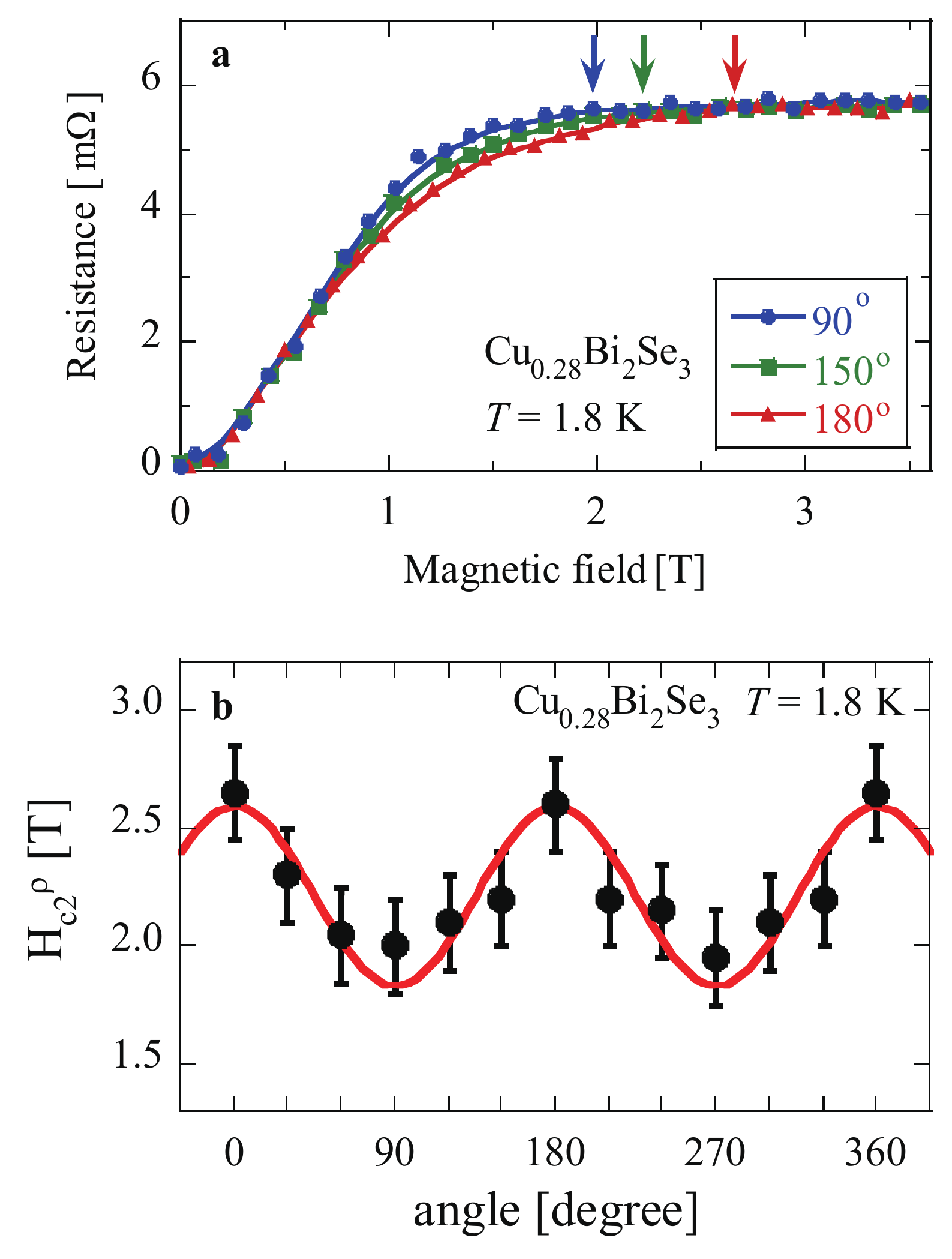}
	\caption{\label{RvsH}
		In-plane upper critical field anisotropy determined by magneto-resistance.
		{\bf a}. Electrical resistance at $T$=1.8 K as a function of the magnetic field for three representative in-plane angles. The arrows  indicate \hct\, 
		which is   defined as the field below which the data point deviates from a straight line drawn from the normal state.
		{\bf b}. Upper critical field $H_{\rm c2}^\rho$ as defined in {\bf a}   is plotted as a function of in-plane angle. The error bar was estimated by assuming that the uncertainty equals twice point-interval in measuring  the resistance {\it vs} field curve.
		The red curve is a sine function.
	}
\end{figure}

Such two-fold symmetry is also seen in the magnetoresistance measurements, as demonstrated in Fig. \ref{RvsAngle} where the electrical resistance under a field of 0.9 T and at $T$=1.8 K is plotted as a function of angle. 
We checked the angle-resolved resistance at $\mu_0H$=3 T and $T$=1.8 K where the sample is  in the normal state and  found only a random noise with an amplitude no bigger than 0.1 m$\Omega$, thus confirming that such  two-fold oscillation in the resistance shown in Fig. \ref{RvsAngle}b is caused by the anisotropy of \hct\ . 
Fig. \ref{RvsH}a shows the resistance as a function of magnetic field at $T$=1.8 K for three representative angles. Similar to the results shown in Fig. \ref{acchi}a, a clear angle dependence is found.
 Fig. \ref{RvsH}b shows the angle dependence of $H_{\rm c2}^\rho$ obtained from the magnetoresistance data.
 A two-fold symmetry is also clearly seen. This result is in agreement  with that seen in the ac-$\chi$ data shown in   Fig. \ref{acchi}b, although such-defined $H_{\rm c2}^\rho$ is slightly higher but is not surprising for its definition. 
 An in-plane anisotropy of  \hct\ is also resolved from the data of electrical resistance {\it vs} temperature under different magnetic fields (see Supplementary Figure 1 and 2).

\begin{figure}[htbp]
	\includegraphics[clip,width=70mm]{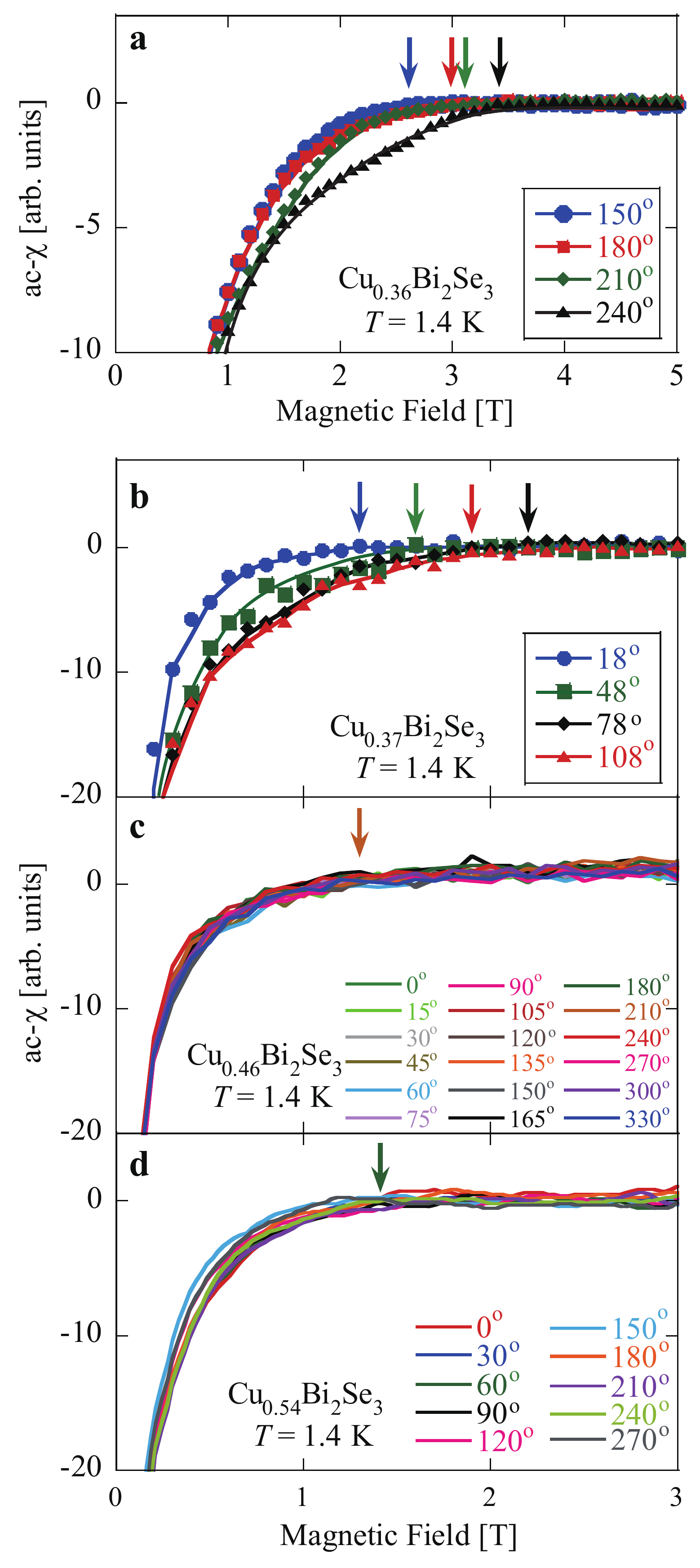}
	\caption{\label{acchi_2}
		Magnetic field dependence of the ac-$\chi$ at different angle $\phi$ for $x$=0.36, 0.37, 0.46 and 0.54.  
		The solid curves  are guides to the eyes. 
		For  sake of clarity, the data points for $x$=0.46 and 0.54 are represented by symbols with  minimized size.
		The arrows  indicate \hct\  defined in the same way as previous figures. For $x$=0.46 and 0.54, the arrow is for $\phi$=15 and 90 degree, respectively.
	}
\end{figure}

Quite often, extracting \hct\ from the magnetic susceptibility has several advantages over that from magnetoresistance measurements. Firstly, the magnetic susceptibility is more sensitive to superconducting volume fraction rather than surface. Secondly, in two-dimensional or layered superconductors, \hct\ determined by resistivity measurements is often inaccurate because of vortex lattice melting \cite{Bonn}. For example, in high-$T_{\rm c}$ cuprates, even in the resistive state, one is still in a regime dominated by Cooper pairings in the presence of vortices \cite{Bonn}.
 For these reasons and the technical merit that ac-$\chi$ can be measured to a lower temperature in our case, below we  discuss the evolution of  gap symmetry based on the ac-$\chi$ data.
 %
Data for $x$=0.36, 0.37, 0.46 and 0.54 
are shown in Fig.\ref{acchi_2}. For $x$=0.36 and 0.37, clear angle dependence was found as in $x$=0.28. In contrast,
the tendency is completely different for samples with larger $x$. 
There is  no angle dependence  in ac-$\chi$   for the samples with $x$ = 0.46 and 0.54. 

\begin{figure}[htbp]
	\includegraphics[clip,width=80mm]{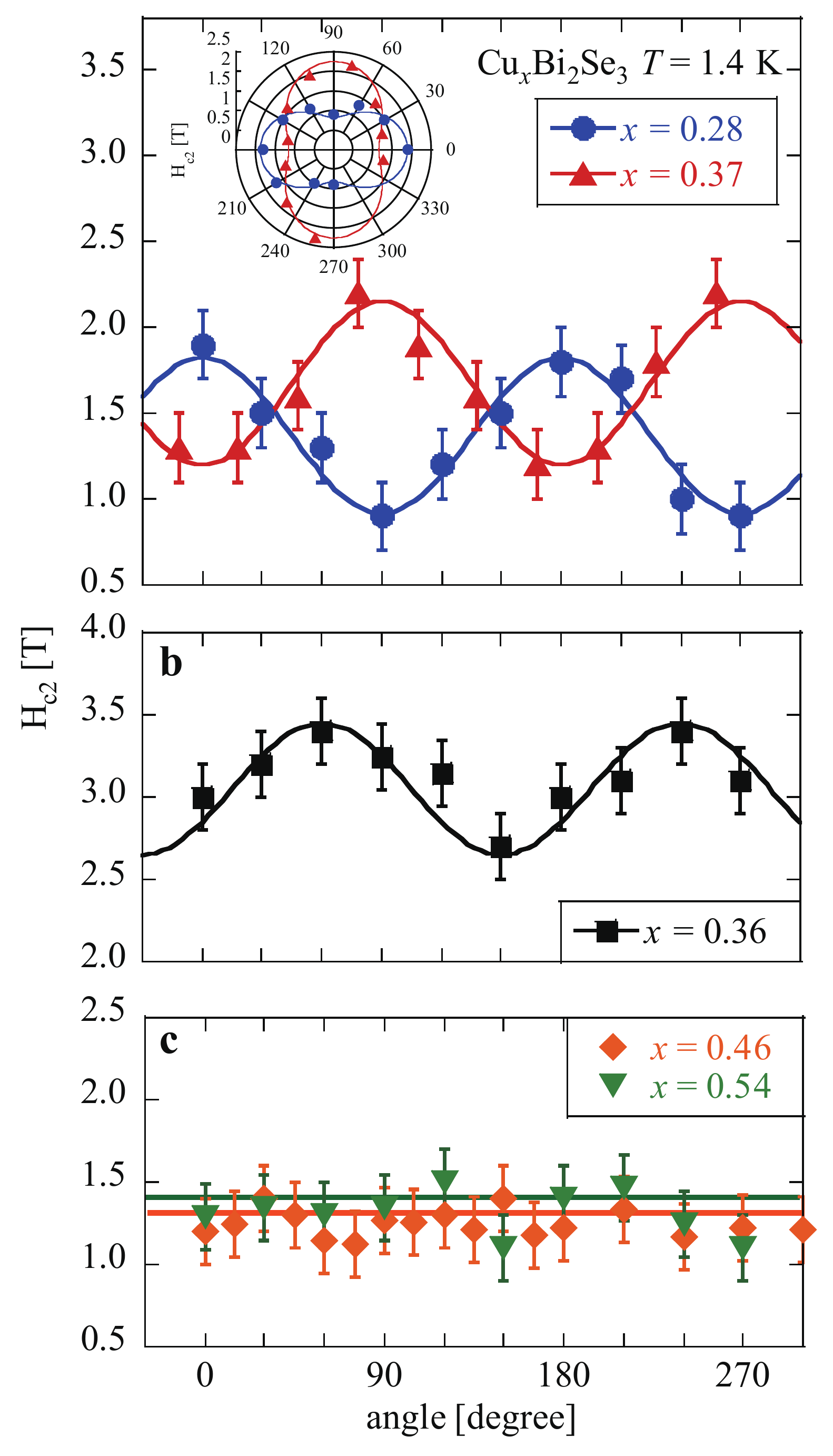}
	\caption{\label{hc2}
		In-plane \hct\  at $T$ = 1.4 K extracted from ac-$\chi$ as a function of angle $\phi$ for all samples. The error bar was estimated by assuming that the uncertainty equals twice point-interval in measuring  the ac-$\chi$ {\it vs} field curve.
		The solid curves are   sine functions. The inset to {\bf a} is a polar plot of the angular dependence \hct\  for $x$=0.28 and 0.37. 
		}
\end{figure}

The angular dependence of \hct\ at 1.4 K determined from the ac-$\chi$  for all samples is plotted in Fig. \ref{hc2} (for more details, see Supplementary Figure 3).
For the samples with smaller $x$ (Fig. \ref{hc2}a, b),  \hct\ shows a large and two-fold anisotropy.
In contrast, for larger $x$ (Fig. \ref{hc2}c), no anisotropy is observed.
The  oscillation amplitude of \hct\ is similar for $x$ = 0.28, 0.36 and 0.37. The magnitude of \hct\ is also similar between $x$ = 0.28  and 0.37, but is larger for $x$=0.36. The origin of this difference is unknown at the moment.

Interestingly, the angle at which \hct\ becomes minimal is 90 and 150 degrees (perpendicular to Se-Se bond) for $x$ = 0.28 and 0.36, respectively, 
but is 0 degree  (along Se-Se bond) for $x$ = 0.37. 
For this crystal structure, the equivalent crystal-axis direction appears every 60 degrees.
If two oscillation patterns have a phase  difference  of 
60 degrees, it can be said that they  are the same crystallographically.
Therefore, the samples with  $x$ = 0.28 and 0.36 have the same gap symmetry.
%
However, 90-degrees difference means that the gap symmetry 
is 
different. 
It is noted that the angle at which \hct\ is minimal corresponds to the direction of  \dv\cite{MatanoKrienerSegawaEtAl2016}.
Therefore, our result indicates that the \dv\ direction  differs for each sample in \cx\   
 even though the carrier density is the same or very similar.
The $x$ = 0.37 sample has the same symmetry as the previous  sample used for NMR measurement 
where
the \dv\ is pinned to Se-Se direction\cite{MatanoKrienerSegawaEtAl2016}, while
 the  $x$ = 0.28 and 0.36 samples have the same symmetry as the sample  used for  specific heat measurements\cite{Yonezawa_Natphys}.
%

\begin{figure}[htbp]
	\includegraphics[clip,width=70mm]{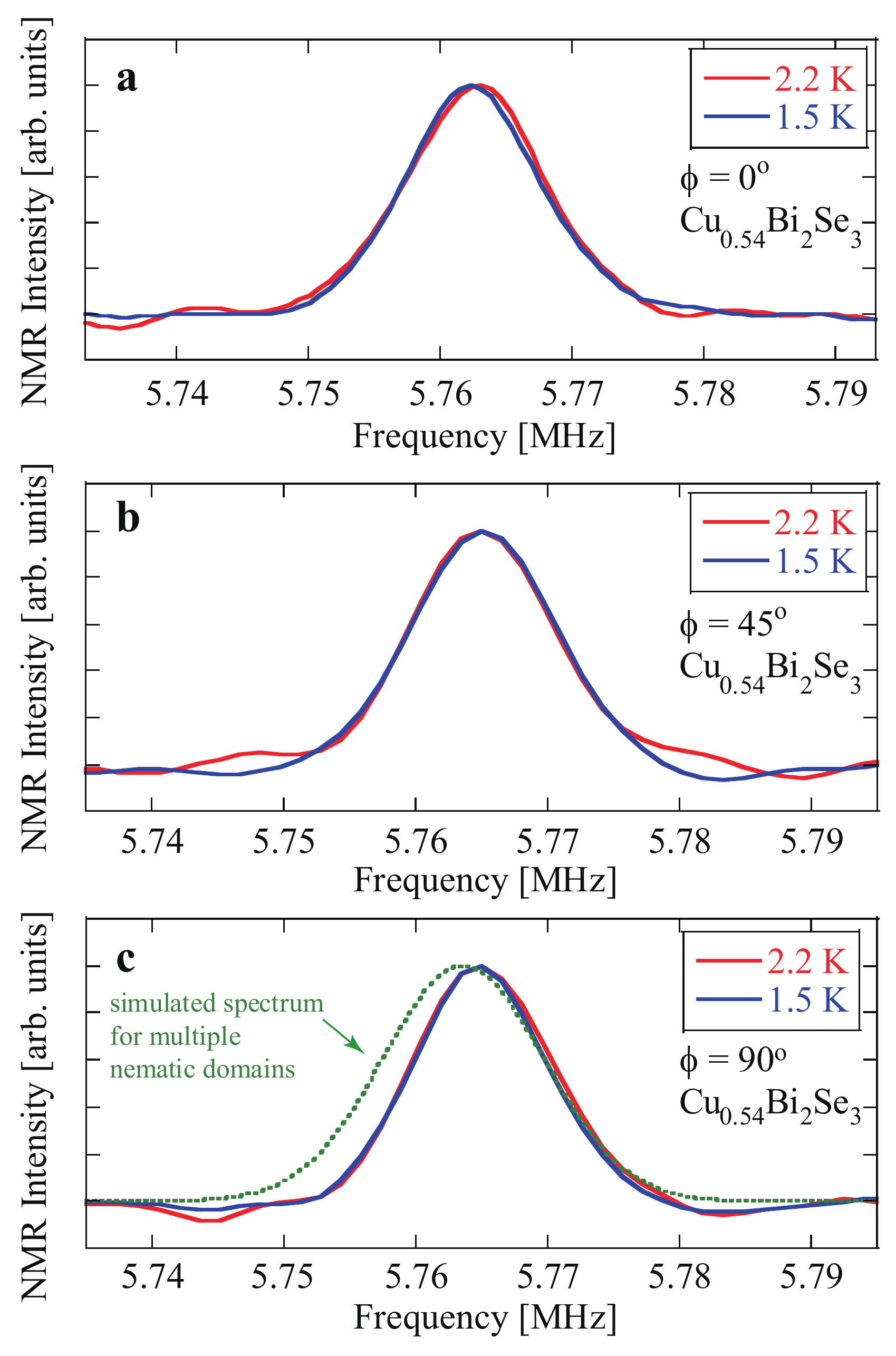}
	\caption{\label{0.54spec}
		NMR spectra and simulation.
		NMR spectra at  $T_{\rm c}$ ($H$=0.7 T) = 2.2 K and below for $x$=0.54 with different angles between $H$ and the $a$-axis ($\bf{a-c}$).  In $\bf{c}$ is shown a  simulation for the spectrum in  presence  of multiple nematic domains, simply assuming that various domains are distributed randomly.
	}
\end{figure}
\textbf{No multiple domains.}
As for the disappearance of  oscillation in  \hct\ for $x$=0.46 and 0.54, a possibility of multiple domains each of which is a nematic phase with \dv\  pointing to a different direction, can be ruled out. In that case, the NMR spectrum for $T<T_{\rm c}$ would be broadened compared to that at $T>T_{\rm c}$. This is because the spectrum coming from the domains with $H\parallel$ \dv\ will shift to a lower frequency \cite{MatanoKrienerSegawaEtAl2016}, but that from domains with $H\perp$ \dv\ will not. 
However, our result (Fig. \ref{0.54spec}  $\bf{a-c}$ ) 
shows no such broadening which is not compatible with the multiple-domains scenario. 
In Fig. \ref{0.54spec}$\bf{c}$, 
we show a simulation for the spectrum in case of multiple domains,  assuming that various domains are distributed randomly as to wipe out oscillations in the angle dependence of \hct. The simulation also assumed that, for $H\parallel$ \dv\ , 
the  spin Knight shift ($K_{\rm s}$) below $T_{\rm c}$ follows the same temperature dependence 
as found in Cu$_{0.3}$Bi$_{2}$Se$_{3}$ \cite{MatanoKrienerSegawaEtAl2016};  
  $K_{\rm s}$=0.046\% and the reduction of $K_{\rm s}$ at $T$=1.5K,  $\Delta K_{\rm s}$=0.043\%, were used in the simulation.

\section{Discussion} 
Fu pointed out that the spin rotation-symmetry breaking (the in-plane Knight shift nematicity) can be understood if the pairing
function  is a doublet representation \cite{Fu_CuxBi2Se3_PhysRevB.90.100509}, $\Delta_{4x}$ or $\Delta_{4y}$, both of which being $p$-wave. For $\Delta_{4x}$ state,   the \dv\ is along the principle crystal axis while it is  orthogonal to the principle crystal axis for $\Delta_{4y}$ state.
%
In the $\Delta_{4x}$ state, there are two point nodes in the superconducting gap, which are 
in the direction perpendicular to the Se-Se bond.
In the  $\Delta_{4y}$ state, there is no node but a minimum in the superconducting gap,  
and the \dv\ is oriented  perpendicular to the gap-minimum direction.
Applying Fu's theory to our results, the $x$ = 0.37 and the previous $x$=0.3 \cite{MatanoKrienerSegawaEtAl2016} samples would correspond to $\Delta_{4x}$, and the samples with $x$ = 0.28 and 0.36 would correspond to $\Delta_{4y}$ state.

As mentioned in Ref. \onlinecite{Fu_CuxBi2Se3_PhysRevB.90.100509}, the theory does not take crystalline anisotropy into consideration so that  $\Delta_{4x}$ and $\Delta_{4y}$ are degenerate.
In real crystals, however, crystalline distortions due to dopants exist, and it was suggested that the dopant position   is important for superconductivity\cite{quench_condiction_PhysRevB.91.144506,Sr_position_PhysRevMaterials.2.014201}.
Hexagonal distortion was indeed reported in \sx\cite{Kuntsevich_distortion}.
The same can be expected in \cx \cite{Yip}. 
Moreover,  strains induced by  quenching  can vary from one sample to another. The chemical processes to obtain a sample with high SF  are complex which includes a quenching process. Only those quenched from a narrow temperature ($\sim$ 560$^\circ$C) show high SF, which suggests that it is important to seize a metallurgically meta-stable phase to obtain the superconductivity. However, the quenching process is less controllable so that strain induced during this process is random among samples.
We believe that    \dv\ pointing to different directions in samples with  the same carrier density  is  due to a different local structural environment such as   strain caused by quenching, dopant-induced  crystal distortion, etc.

The most intriguing and surprising finding of this work is that 
 the oscillation in \hct\ disappears
for the samples with  $x$= 0.46 and 0.54, as seen in Fig. \ref{hc2}{\bf b}. 
Judging from 
the NMR spectrum width which is almost the same for $x$=0.36, 0.48 and 0.54 in particular, we conclude that the sample homogeneity  is very similar among them. Also,  a multiple-domains scenario can be ruled out as discussed in the previous section. 
We interpret the surprising evolution of \hct\  as due to an emergent fully-opened isotropic gap for large $x$.
It was previously pointed out that the odd-parity superconductivity with two-component $E_u$ representation admits two possible phases,  nematic and chiral \cite{chiral_PhysRevB.94.180504, WangMartin}. Under some circumstances, a chiral, full-gap, state becomes more stable as compared to a nematic state. One of the crucial parameters important for selecting between the two states  is  the  Fermi surface shape. It has been theoretically shown by several groups that  the chiral state can be stabilized when the Fermi surface becomes more two-dimensional \cite{mizushima_2018arXiv180906989U,Chiral_PRB,WangQH}.
Experimentally,  photoemission and quantum oscillation  measurements have  suggested that the Fermi surface becomes more two-dimensional as carrier density increases  \cite{Arpes,LiLu}. 
Indeed, the Knight shift result indicates that carriers increased in the samples of $x$=0.46 and 0.54 as compared to $x\leq$0.37.
Therefore, the dramatic change in the angle dependence of the \hct\  can be naturally explained as due to a gap symmetry change from  nematic  to  chiral.
The lack of change of the Knight shift below $T_{\rm c}$ for all angles (Fig.\ref{0.54spec}) as contrary to what found in the nematic phase (\cite{MatanoKrienerSegawaEtAl2016}) is consistent with existing theory that the \dv \
for the chiral phase would be pointing to the $c$-axis \cite{mizushima_2018arXiv180906989U}. Thus, our finding suggests that superconductors  with  strong spin-orbital interaction \cite{Fu_Berg_PhysRevLett.105.097001} is a land much more fertile  than we thought.

In summary, single crystal samples of  \cx\ with unprecedented high doping levels were newly synthesized and investigated.
By NMR measurements, we found for the first time that the carrier density increased further by Cu doping beyond $x$=0.37.
By magnetic susceptibility measurements, we found that the in-plane \hct\ shows a clear two-fold oscillation for the samples with $x$ = 0.28, 0.36 and 0.37 which have similar carrier density as evidenced by the Knight shift.
However, the angle at which \hct\ becomes minimal is different by 90 degrees 
between different samples. This indicates that the  direction of the \dv\ is different from crystal to crystal
due to a different local structure caused by  strain during the quenching process,  dopant-induced crystal distortion, etc.. 
In the samples with $x$ = 0.46 and 0.54, the 
two-fold oscillation is suppressed, which indicates a gap symmetry change from nematic to isotropic as carrier density increases. These findings enriched the contents of topological superconductivity in doped Bi$_2$Se$_3$, and we hope that our work will stimulate further studies on possibly even more exotic superconducting state (possible chiral state) in the high-doped region of \cx\  as well as on bulk topological superconductors in general.

\section{Methods}
\noindent
\textbf{Single crystal growth and characterization.}
Single crystals of \cx\ were prepared by intercalating Cu into \BS\ by the electrochemical doping method described in Ref. \onlinecite{Kriener_PhysRevB.84.054513}.
First, single crystals of \BS\ were grown by melting stoichiometric mixtures of 
elemental Bi (99.9999\%) and Se (99.999\%) at 850 $^{\rm o}$C for 48 hours in sealed evacuated quartz tubes. 
After melting, the sample was slowly cooled down to 550 $^{\rm o}$C over 48 hours
and kept at the same temperature for 24 hours.
Those melt-grown \BS\ single crystals were cleaved into smaller rectangular pieces of about 14 mg.
They were wound by bare copper wire (dia. 0.05 mm), and used as a working electrode.
A Cu wire with diameter of 0.5 mm  was used both as the counter (CE) and the reference electrode (RE).
We applied a current of 10 $\mu$A in a saturated solution of CuI powder (99.99\%) in acetonitrile (CH$_3$CN).
The obtained crystals samples were then annealed at 560 $^{\rm o}$C for 1 hour in sealed evacuated quartz tubes, and quenched into water.
After quenching, the samples were covered with epoxy (STYCAST 1266) to  avoid deterioration.
The Cu concentration $x$ was determined from the mass increment of the samples.
To check the superconducting properties, dc susceptibility measurements were performed using a superconducting quantum interference device (SQUID) with
the vibrating sample magnetometer (VSM).
 
\noindent
\textbf{NMR measurements.}
The $^{77}$Se-NMR spectra were obtained by the fast Fourier
transformation of the spin-echo  at a field of $H_0 $ = 1.5 T. 
The Knight shift $K$ was calculated using nuclear gyrometric ratio $\gamma_{\rm N}$ = 8.118 MHz/T for $^{77}$Se.

\noindent
\textbf{Angle-resolved \hct\  measurements.}
\hct\  
was determined from
ac susceptibility  by measuring the inductance of  in-situ NMR coil. 
Angle-dependent measurements 
were performed by using a piezo-driven rotator (Attocube  ANR51) equipped with Hall  sensors  to determine the angle between  magnetic field and  crystal axis. 
The ac-$\chi$ vs $H$ data in the normal state were fitted by a linear function (a constant line). \hct\  was defined as a point off the straight line. A typical example is shown in Supplementary Information.

\noindent
\textbf{Magnetoresistance measurements}. The angle-dependent electrical resistance was measured by the standard four-electrode method in a Physical Properties Measurement System (PPMS, Quantum Design) with a mechanical rotating probe. The 
building of  electrodes were carried out in a glove box filled with high-purity Ar gas to prevent sample from degradation. 
The electrodes were made such that the current direction is 
along the $a$-axis. 
The excitation currents are 0.1-1 mA to make a compromise of the Joule heating and the measurement accuracy. 

\section{Acknowledgments}
We thank Y. Inada for help in Laue diffraction measurements and S. Kawasaki for help in some of the \hct\  measurements, Markus Kriener and T. Mizushima  for useful discussions. 
This work was supported in part by the JSPS/MEXT  Grants (Nos. JP15H05852, JP16H04016, and JP17K14340) and  MOST grant 
No. 2017YFA0302904.
 \section{Authors contributions}
G.-q.Z planned and supervised the project. T.Kawai, Y.K and T.Kambe synthesized  the  single crystals. 
T.Kawai, Y.K, Y.H and K.M  performed magnetic susceptibility,    Y.H and K.M conducted NMR, and C.G.W performed magneto-resistance measurements. G.-q.Z wrote the manuscript with inputs from  K.M.  All authors discussed the results and interpretation.  

\section{Competing financial interests}
The authors declare no competing  interests.

\section{Materials and Correspondence}
Correspondence and requests for materials should be addressed to G.-q.Z.

\section{Data availability}
The data that support the findings of this study are available on reasonable request.

\end{document}